# Gate-controlled neuromorphic functional transition in an electrochemical graphene transistor

Chenglin Yu[1,†], Shaorui Li[1,†], Zhoujie Pan[6,†], Yanming Liu[7,8], Yongchao Wang[1,2], Siyi Zhou[1], Zhiting Gao[1,2], He Tian[7,8], Kaili Jiang[1,3], Yayu Wang[1,4,5], Jinsong Zhang[1,4,5*]

[1]*State Key Laboratory of Low Dimensional Quantum Physics, Department of Physics, Tsinghua University, Beijing 100084, China*

[2]*Beijing Innovation Center for Future Chips, Tsinghua University, Beijing, China*

[3]*Tsinghua-Foxconn Nanotechnology Research Center, Department of Physics, Tsinghua University, Beijing 100084, China*

[4]*Frontier Science Center for Quantum Information, Beijing, 100084, China*

[5]*Hefei National Laboratory, Hefei 230088, China*

[6]*XingJian College, Tsinghua University, Beijing 100084, China*

[7]*School of Integrated Circuits, Tsinghua University, Beijing 100084, China*

[8]*Beijing National Research Center for Information Science and Technology (BNRist), Tsinghua University, Beijing 100084, China*

[†] These authors contributed equally to this work.* Email: <u>jinsongzhang@tsinghua.edu.cn</u>

**Abstract**

Neuromorphic devices have gained significant attention as potential building blocks for the next generation of computing technologies owing to their ability to emulate the functionalities of biological nervous systems. The essential components in artificial neural network such as synapses and neurons are predominantly implemented by dedicated devices with specific functionalities. In this work, we present a gate-controlled transition of neuromorphic functions between artificial neurons and synapses in monolayer graphene transistors that can be employed as memtransistors or synaptic transistors as required. By harnessing the reliability of reversible electrochemical reactions between C atoms and hydrogen ions, the electric conductivity of graphene transistors can be effectively manipulated, resulting in high on/off resistance ratio, well-defined set/reset voltage, and prolonged retention time. Overall, the on-demand switching of neuromorphic functions in a single graphene transistor provides a promising opportunity to develop adaptive neural networks for the upcoming era of artificial intelligence and machine learning.

**Main text**

In the current era of rapidly escalating computational demands, the imminent cessation of Moore's Law[1–3] and the von Neumann bottleneck constitute impediments to the further enhancement of computational efficiency.[4,5] To address the limitations of the traditional paradigm, researchers have turned to neuromorphic systems, which replicate the neural functionality of the human brain. Achieving practical neuromorphic systems requires a deep understanding of brain functions, as well as the development of innovative electronic devices that can mimic neural circuits. One promising electronic component for this purpose is the memristor[6,7], which offers the potential for high-density memory and neuromorphic computing.[8,9]

Based on retention time, memristors are classified into volatile and non-volatile types, the former of which can be used as artificial neurons while the latter can function as artificial synapses.[10–12] Currently, the majority of research on memristive devices centers on the analysis and improvement of their performance, specifically in relation to aspects such as the conversion between short-term and long-term plasticity[13–15], precise neurocomputational profiling[16,17], and the integration and transmission of signals in simulated neurons[18–20]. However, these devices are all operated within the confines of either volatile or non-volatile memristors, which greatly restricts the versatility and superiority of artificial neural networks. Consequently, there is a high demand for the discovery of an adaptable memristor with switchable volatile and non-volatile functionalities. The capability to dynamically reconfigure the network and effectively execute diverse tasks holds considerable significance for the development of intelligent biorealistic computers in the future.[4,21] Recently, research groups have reported achievements in the field of volatile and non-volatile switching transitions within devices based on materials such as $CuInSe_2$, accomplished by employing different compliance currents.[22–24] However, it should be noted that the underlying switching mechanisms in these devices are still contingent upon the formation and destruction of Cu conductive filaments. Due to the inherent instability associated with ion migration, the stability or retention time remains to be improved.[25,26]

To address these issues, we investigate the use of monolayer graphene, an 2D material with high electron mobility and chemical stability[27], as the channel material of the memristive devices, which can provide an open architecture for multi-terminal configurations with compatible gate tunability. While graphene has been extensively studied as an electrode material or utilizing graphene protonation for memristors in memory device applications[28–30], its potential as the critical switching material has not been fully explored. In this work, we present a unique neuromorphic device based on an electrochemical graphene transistor, where the functional transition between artificial synapses and artificial neurons can be efficiently controlled through applying different gate voltages, which is more versatile and repeatable compared to other reconfigurable devices controlled by specific electric pulses[21] or compliance currents.[22,31,32] Moreover, in our devices the inherent electrochemical reactions between carbon atoms and hydrogen ions guarantee the uniformity of device performance, which offers a promising solution for the development of high-precision memristor-based computational systems.

The device structure of our electrochemical graphene transistors (EGTs) is depicted in Figure 1a. The graphene channel is exposed to the hydrogen ion electrolyte (HIE) comprised of liquid organic solvent and dissolved $H^+$ ions (see Methods section for details). The application of gate voltage ($V_G$) between the gate terminal and source electrode regulates the migration of $H^+$ ions towards or away from the interface, leading to a significant alteration in the resistance of the graphene channel. This phenomenon underscores the potential of graphene transistors in transducing small changes in voltage or electrochemical reactions into measurable shifts in electrical response.

Figure 1b illustrates the drain-source current ($I_{DS}$) of our EGTs under the gate control. When $V_G$ is restricted within the range of -1.0 V to 1.5 V, the characteristic V-shaped behavior in this curve originates from the linear dispersion of relativistic Dirac fermions in pristine graphene[33,34], defined as low resistance state (LRS) in our experiments. However, as gate voltage exceeds the hydrogenation voltage ($V_H$), which is around 1.8 V, the $H^+$ ions accumulating on the top of the graphene surface start to react with the graphene, causing a change in the hybridization of the carbon atom from $sp^2$ to $sp^3$, which then localizes the conduction electrons and switches to the high resistance state (HRS).[35–38] This results in a sharp drop of the $I_{DS}$ to nearly zero. When $V_G$ returns to -1 V, the graphene is dehydrogenated, and $I_{DS}$ promptly recovers back to the LRS. Notably, our previous work has demonstrated that such on/off ratio can reach $10^8$ with high reversibility for up to 1 million switching cycles[37,38], which presents exciting implications for the development of advanced devices that require high stability and specificity through the delicate control of C-H bond concentration in graphene.

To further investigate hydrogenation under the control of drain-source voltage

($V_{DS}$), we fix $V_G$ while sweeping $V_{DS}$ to switch EGTs between HRS and LRS. As shown in Figure 1c and d, the I-V loops exhibit diverse switching behaviors at different $V_G$. When the EGT is set to LRS with $V_G$ in the range of -1.0 V to 0.2 V, the I-V curves show no hysteresis and the EGT behaves as a resistor (Figure S1). In contrast, when $V_G$ is fixed at the values within the range of 0.4 V to 1.6 V, lower than $V_H$, the I-V characteristics demonstrate a splendid non-volatile memristive behavior, in which the resistance can be alternately switched to the LRS (SET process) at positive SET voltage ($V_{SET}$) and recovered to the HRS (RESET process) at negative RESET voltage ($V_{RESET}$). Moreover, both the $V_{SET}$ and $V_{RESET}$ exhibit a systematic gate dependence with the switching ratio of $R_{HRS}/R_{LRS}$ up to $10^6$. Furthermore, the I-V loops obtained during cyclic sweep of $V_{DS}$ present desirable consistency and stability at different $V_G$, as illustrated in Figure S2.

The underlying mechanism can be attributed to the reversible hydrogenation reaction between graphene and $H^+$ ions under the combined control of $V_G$ and $V_{DS}$. For instance, when $V_G$ is set to 1.6 V and the graphene channel is in HRS, the effective gate voltage ($V_{G,Eff}$) applied to the graphene segment near the drain electrode would reduce upon increasing the $V_{DS}$. The $V_{G,Eff}$ close to the drain electrode can be roughly calculated by using $V_{G,Eff} = V_G - V_{DS}$, if the contact resistance is negligible. Once the $V_{G,Eff}$ drops below the dehydrogenation voltage ($V_{DH} \sim 0$ V), this graphene segment would be dehydrogenated and then SET the graphene channel to LRS. This is a non-volatile state when $V_{DS}$ returns to zero because the applied $V_G$ is lower than $V_H$ and higher than $V_{DH}$, thereby preventing both hydrogenation and dehydrogenation reactions. Conversely, decreasing the $V_{DS}$ would increase the $V_{G,Eff}$ of the graphene segment close to the drain electrode until it exceeds the $V_H$, triggering the successive hydrogenation reaction of the graphene channel and RESET it to HRS. Figure S3 explicitly displays the spatial variation of $V_{G,Eff}$ under varying $V_{DS}$. Since the $V_{SET}$ and $V_{RESET}$ are determined by the critical $V_{DS}$ when $V_{G,Eff}$ of the graphene segment next to the drain electrode matches $V_{DH}$ or $V_H$, different $V_G$ generates different values of $V_{SET}$ and $V_{RESET}$. When $V_G$ is kept at a voltage higher than $V_H$ (Figure 1d), the EGT can be switched alternatively between

HRS and LRS by sweeping $V_{DS}$. However, it becomes volatile and stays in HRS when $V_{DS}$ is retracted to zero due to the hydrogenation reaction with $V_G > V_H$. This versatile gate-controlled evolution of volatile and non-volatile memristive behaviors enables us to operate the EGT as a multi-functional device in the neuromorphic network.

To substantiate the aforementioned switching mechanism, Raman measurements were conducted. The Raman spectra of monolayer graphene are characterized by a prominent G peak, which appears near 1,580 cm$^{-1}$ and originates from in-plane vibrational ($E_{2g}$) mode scattering.[39,40] Notably, the hydrogenation reaction of graphene lattice can lead to the emergence of the D and D' peaks at approximately 1,340 cm$^{-1}$ and 1,600 cm$^{-1}$, respectively. Because the activation of D peak necessitates the presence of a defect via an intervalley double-resonance Raman process[39,40], the large D-peak singles reflect the strong breaking of translational symmetry of C-C $sp^2$ bonds and the formation of massive C-H $sp^3$ bonds[41], behaving like a great number of defects in the graphene lattice. Therefore, the intensity of the D peak can serve as an indicator of the level of defect concentration and the resistivity within the graphene channel.

Figure 2 presents the evolution of Raman spectra in a complete I-V loop with $V_G$ fixed at 1.4 V throughout the experiment. At $V_{DS} = 0$ and the EGT is set to the HRS, the entire graphene channel displays strong D-peak signals. When $V_{DS}$ is elevated to 0.5 V, the graphene channel remains insulating owing to the fact that the $V_{G,Eff}$ at the drain region is still higher than $V_{DH}$. At $V_{DS} = 1.5$ V, the D peak initiates disappearance first in the vicinity of the drain region while the graphene channel adjacent to the source region continues to exhibit moderate D-peak intensity, which proves that the dehydrogenation starts from the drain electrode in the SET process. Subsequently, when $V_{DS}$ increases to 2.5 V, the entire graphene channel undergoes sufficient dehydrogenation reaction and reverts to the LRS. Moreover, when negative $V_{DS} = -0.4$ V is applied to the drain electrode, the $V_{G,Eff}$ of the graphene segment around there is approaching $V_H$ and the hydrogenation reaction happens near the drain area. However, we still observed plenty of D-peak signals near the source electrode, which probably arises from the presence of residual C-H bonds even when $V_{DS}$ is set to 2.5 V (Figure

S5). These active C-H bonds would facilitate the hydrogenation reaction of neighboring C atoms[36,42], thus creating more D-peak signals when the $V_{G,Eff}$ is slightly increased and close to $V_H$. As the decrease of $V_{DS}$ to − 0.8 V, the whole graphene channel is RESET to the HRS once again. Correspondingly, Raman spectra gathered at the marker position of the graphene channel are presented in Figure 2a, demonstrating the alterations of the D-peak intensity at different $V_{DS}$. More Raman measurements in a different device exhibit the same evolution behaviors as shown in Figure S6. Figure 2c presents a schematic diagram that effectively showcases the evolution of the graphene lattice as it reacts with $H^+$ ions throughout an entire switching cycle. For this purpose, we employ the chair conformation of hydrogenated graphene, which has been proposed as the most stable and favorable structure amongst all conformers.[35,36]

Figure 3a depicts the resistance values of the LRS and HRS, as well as the on/off ratio of our EGTs at varying $V_G$, exhibiting a negligible degree of variation as the $V_G$ is increased. These results highlight the stable electrical characteristics of our EGTs under different $V_G$ levels. The LRS is characterized by a minimum resistance of only a few kilo-ohms, attributable to the exceptional electrical conductivity of graphene, while its HRS is in the giga-ohm range, as a result of the insulating properties of hydrogenated graphene with a large band gap.[37] It should be noted that the reported resistance of HRS is only a conservative estimate because of the presence of leak current between the gate and drain electrode (Figure S7), As a result, the actual on/off ratio of the EGTs should be considerably higher than the estimated value of ~ $10^6$.

In contrast to the nearly constant on/off ratio, the $V_{SET}$ and $V_{RESET}$ exhibit linear dependence on $V_G$ as presented in Figure 3b. More specifically, the $V_{SET}$ and $V_{RESET}$ are related to the critical $V_{DS}$, at which the $V_{G,Eff}$ of the graphene segment adjacent to drain electrode equates with $V_{DH}$ and $V_H$ respectively. This $V_{G,Eff}$ can be approximated by utilizing the formula $V_{G,Eff} \approx V_G - V_{DS}$, thereby yielding the $V_{SET} \approx V_G - V_{DH}$ and $V_{RESET} \approx V_G - V_H$. Unlike other types of memristive devices, such as those based on ionic migration[7,43,44], phase changing[45,46], or electron spin[47], the switching mechanism of our EGTs is directly determined by inherent voltage-induced electrochemical reaction. Although traditional memristors utilizing ionic migration have shown good switching capabilities, the stochasticity associated with the kinetics of ion movement leads to a strong variability during cyclic switching.[26,48] Whereas, our EGTs harness the well-defined electrochemical reaction voltage to improve switching accuracy, as evidenced in Figure 3c-d by the consistent switching voltages with narrow distribution over 100 consecutive cycles. The representative I-V loops for volatile and non-volatile memristive behaviors are displayed in Figure S8. This unique property offers accurate analog programming capabilities, which are particularly desirable for neuromorphic computing.

Synapse are fundamental in neuroscience, shaping neural connections for complex cognition. Here, we simulate the behaviors of artificial synapses within the non-volatile switching mode in our EGTs. To ensure the non-volatility, the $V_G$ is fixed at 1 V. As shown in Figure 4a, the cyclic sweeping of $V_{DS}$ in the range of 0 to 1.4 V gradually increases the $I_{DS}$ and switches HRS to LRS, which results from the progressive dehydrogenation reaction of the graphene channel when $V_{DS}$ exceeds $V_{SET}$. The corresponding changes in the conductivity simulate the modification of artificial synaptic weight via input voltages. Similarly, the continuous sweeping of $V_{DS}$ in the range of -0.8 V to 0 V decreases the $I_{DS}$ step by step and finally returns to HRS as displayed in Figure 4b.

Additionally, the conductance of our non-volatile memristor can be dynamically controlled through specific electrical stimuli, as evident from the results presented in Figure 4c. Specifically, a series of consecutive positive voltage pulses result in an incremental increase in memristor conductance, while a series of negative voltage pulses lead to an incremental decrease. Here we have employed a pulse width of 10 ms. However, it should be noted that a pulse width of 1 ms, which more closely approximates the width of biological electrical signals, has also been tested and demonstrated comparable modulation effects (Figure S9).

Figure 4d illustrates the retention time, all the conductance states exhibit negligible variation for 5 minutes, indicating a robust retention capacity. It is plausible to assume that this retention phenomenon may persist for a more extended duration. This is attributed to the stable lattice structure of hydrogenated graphene, reinforced by the control of gate voltage, which maintains the conductive states from external fluctuations.

In addition to the aforementioned functionalities, another significant aspect of our EGTs is their potential to function as synaptic transistors, in which the gate terminal can be utilized to systematically modify the synaptic weight.[14,49] The hysteresis behavior displayed by the transfer curve in Figure 1b enables the modulation of the conductance of the graphene channel in a non-volatile manner by controlling $V_G$. As demonstrated in Figure 4e, the continuous sweep of $V_G$ towards $V_H$ gradually initiates the hydrogenation of the graphene channel, leading to a slow decrease of $I_{DS}$ during each sweeping cycle. Conversely, the gradual increase of $I_{DS}$ can be achieved by the dehydrogenation reaction when sweeping $V_G$ to negative values. To maintain the degree of hydrogenation and obtain coincident $I_{DS}$ of two adjacent sweeping cycles, the starting voltage of each sweeping cycle is set to 0.2 V, where the Fermi level lies at the Dirac point of graphene. Moreover, the $I_{DS}$ can be effectively controlled by applying a series of pulses to the gate terminal as shown in Figure S13. In addition, further characterizations of pulse synaptic plasticity of both artificial synapse mode and synaptic transistor mode are shown in Figure S10-S12.

Furthermore, we also conduct simulations of neuronal firing behavior. The $V_G$ is fixed at 2 V to ensure the volatile switching behavior and mimic the neuron excitation by applying a pulse $V_{DS}$. Figure 5a shows that when the pulse amplitude is higher than $V_{SET}$ and is sustained for a sufficient duration (80 ms), the artificial neuron spikes and returns to its initial state after the excitation. Here, we use a read voltage ($V_{READ}$) of 0.05 V to detect the neuron's status. In the case that the pulse width is inadequate to reach the threshold, the artificial neuron will not fire a spike, as shown in Figure 5b when the pulse width is only 10 ms. Moreover, Figure S14 demonstrates that the excitation threshold can also be attained by increasing the amplitude of the input pulse while keeping the width constant.

In biological neural system, an individual neuron is tasked with the integration and processing of a sequence of neural impulses received from multiple synapses.[19] Figure 5c illustrates the intricate electrophysiology mechanism involved in neural activities. To simplify this dynamic process, a leaky integrate-and-fire (LIF) model with a dynamic threshold is commonly employed, which enables a more accessible description of the complex neuron activities.[19,20,50] In this regard, we investigate the corresponding behaviors of our EGTs in response to consecutive electric pulses. As depicted in Figure 5d, the output current exhibits a state of inactivity initially and subsequently initiates an uptrend after the application of several pulses. Notably, an increase in pulse interval entails a larger number of input pulse to fire a spike. These distinctive properties effectively emulate the behavior envisaged by the LIF model, thus underscoring the exceptional potential of our EGTs in the development of artificial neurons.

In summary, we have successfully developed a unique multi-function neuromorphic device based on the EGTs, which displays a seamless transition between artificial synapse and neuron through the versatile control of $V_G$. It would bridge the gap between modification of synaptic weight and integration of signals for neural spiking and neuromorphic computing. The Raman measurements indicate that the switching mechanism between HRS and LRS is realized by reversible hydrogenation and dehydrogenation reactions of graphene and hydrogen ions under the combined control of $V_{DS}$ and $V_G$. The exceptional performance of our EGTs, including a high on/off resistance ratio, well-defined $V_{SET}/V_{RESET}$, and prolonged retention time, is attributed to the inherent electrochemical reaction, which distinguishes them from traditional memristors based on ion migration. Moreover, the linear dependence of $V_{SET}/V_{RESET}$ on $V_G$ makes our EGTs more suitable for accurate neuromorphic computing. These highly adjustable and reconfigurable characteristics of our EGTs provide more flexibility in tuning synaptic weight and neural spiking, optimizing circuit design, and building an adaptive neural network.

**Method**

**Device preparation.** Graphene monolayer flakes were obtained by using mechanical exfoliation method on 285 nm $SiO_2$/Si wafers. After that, the drain and source electrodes (Cr/Au/AlO$_x$, typical thickness ~3/50/50 nm) were deposited using standard electron-beam lithography and thermal evaporation process. Then a layer of Poly(methyl methacrylate) (PMMA) is spin-coated on the electrode, while only the graphene channel is exposed to the HIE. The contact pads for wire bonding are always protected during AlO$_x$ deposition and PMMA coating. For devices used in electric measurements, the gate electrode was made of a layer Ti/Pt (5/50 nm) coated on a doped silicon wafer. One layer of thermal plastic (~60 μm) was used to seal the top gate electrode and $SiO_2$/Si wafer. For devices used in Raman spectroscopy measurement, a long strip of platinum foil was used as the gate electrode instead. The detailed assembling procedures and electrolyte injection were reported in our previous works[37,38]. The hydrogen ion electrolyte consisted of 0.4 mol/L bis(trifluoromethane)sulfonamide (HTFSI) in liquid polyethylene glycol (PEG) with an average $M_n$ of 600. The PEG was heated to 100 ºC for few hours followed by vacuum evacuation to remove residual moisture.

**Electric measurements.** For the DC measurements, two Keithley 2400 SourceMeters were used to apply $V_{DS}$ between the drain and source electrodes and $V_G$ between the gate and source electrode respectively. The transfer curve is obtained at $V_{DS}$ = 5 mV. For the temporal response measurements, a semiconductor device analyzer (Keysight B1530) was used to apply the pulses of $V_{DS}$ and $V_G$, while the $I_{DS}$ and gate current ($I_G$) were recorded simultaneously. All the experiments were carried out at room temperature (~25 °C).

**Raman spectroscopy measurements.** Raman spectroscopy was measured by a HORIBA LabRAM HR Evolution spectrometer with 532 nm laser excitation and a grating with 600 lines per mm. The background signals contributed by the electrolyte were deducted in the Raman spectra. During the Raman measurements, two Keithley 2400 SourceMeters were connected to the device to control $V_G$ and $V_{DS}$ synchronously.

When the applied $V_{DS}$ reached the target value, it was expeditiously reduced to zero prior to conducting Raman mapping. A step size of 1 µm was used in the Raman mapping.

**Data availability**

The data that support the plots within this paper and other findings of this study are available from the corresponding authors upon reasonable request.

**Supporting information**

Further details on devices description; a plot of I-V curve with negative $V_G$; multiple I-V loops with varied $V_G$; switching mechanism modeling; structures of EGT devices; supplementary Raman mappings; resistance measurements of HRS; modulation of conductance states of artificial synapse, artificial neuron and synaptic transistor mode; excitatory postsynaptic current; paired pulse facilitation in artificial synapse mode; period-dependent synaptic potentiation in synaptic transistor mode; simulation of pattern recognition.

**Acknowledgments:**

We thank Fan Wu, Puyang Cai, and Zhenqi Hao for helpful discussions. This work is supported by the National Key R&D Program of China (grants No. 2018YFA0307100), National Natural Science Foundation of China (grant No. 12274252), the Basic Science Center Project of NSFC (grant No. 52388201), and the Innovation Program for Quantum Science and Technology Grant No. 2021ZD0302502.

**Author contributions**

J.S.Z., Y.Y.W., and K.L.J proposed and supervised the research. C.L.Y., S.R.L, S.Y.Z, and Z.T.G. fabricated the devices and carried out the electric measurements. C.L.Y. and Y.C.W. measured the Raman spectra. Z.J.P., Y.M.L, and H.T did the simulation of pattern recognition. J.S.Z. and C.L.Y. prepared the manuscript with comments from all the authors.

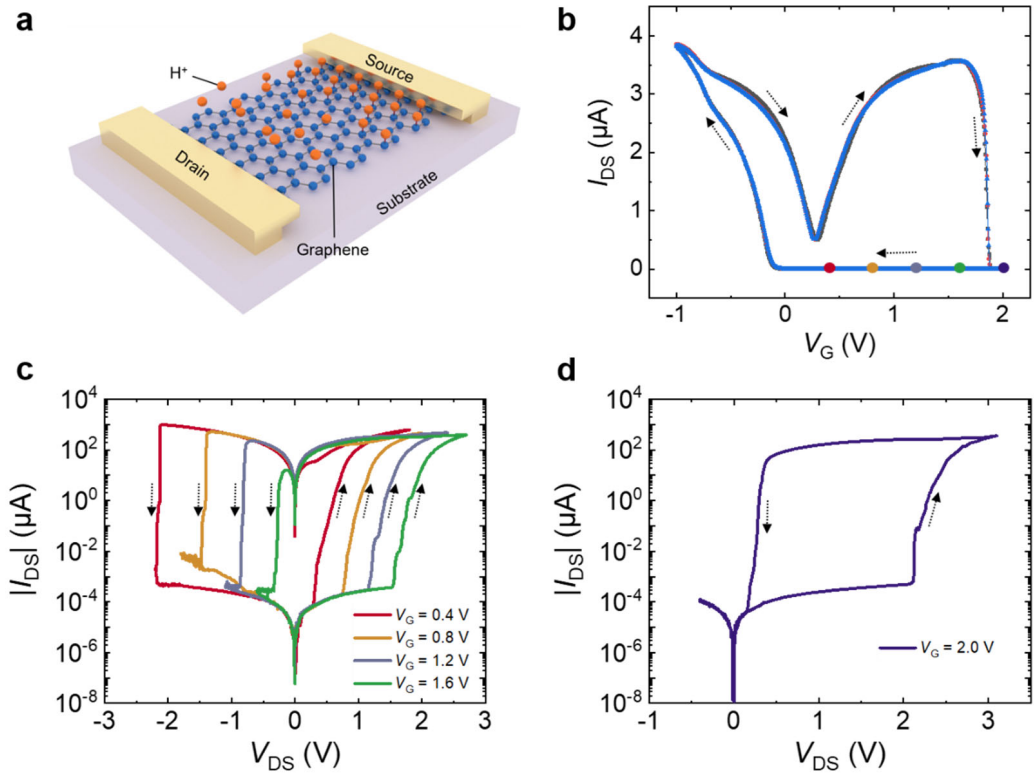

**Figure 1. Gate-controlled memristive behaviors in the EGT.** (a) Schematic of our neuromorphic graphene device. The graphene channel is immersed in the hydrogen ion electrolyte. (b) Three consecutive cycles of transfer curves measured at room temperature (25 °C). The black arrows indicate the sweep directions of $V_G$. The applied $V_{DS}$ is fixed at 5 mV. (c) I-V loops under cyclic sweep of $V_{DS}$ at varied $V_G$. The non-volatile memristive switching behaviors exhibit systematic dependence on $V_G$. (d) Volatile switching behavior with $V_G$ = 2.0 V. The corresponding $V_G$ of the I-V loops in (c) and (d) are marked by the solid dots in (b) with the same colors.

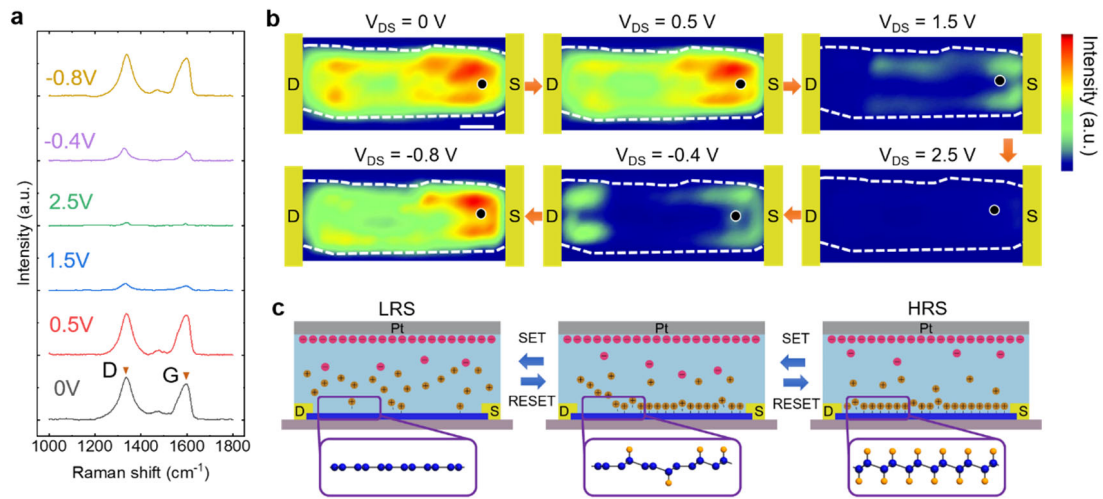

**Figure 2. In-situ Raman characterization of hydrogenated graphene at $V_G$ = 1.4 V.** (a) Raman spectra of hydrogenated graphene at varied $V_{DS}$ in a switching cycle. The applied $V_{DS}$ is labeled next to the curves. (b) Raman mapping of D-peak intensity during $V_{DS}$ sweeps from 0 V to 2.5 V, then returns to -0.8 V. Regions with high D-peak intensity (red) signify high resistivity. The graphene channel is marked by the dashed white line, with source and drain regions on the right and left, respectively. The scale bar is 2 μm. The black spots represent the measuring points for Raman spectra displayed in (a). (c) Schematic representation of hydrogenation reactions between graphene lattice and $H^+$ ions. The arrows indicate the directions of SET and RESET processes during the alternative switching between HRS and LRS. The hypothetic lattice structures of hydrogenated graphene for different degrees of hydrogenation are illustrated in the bottom panels.

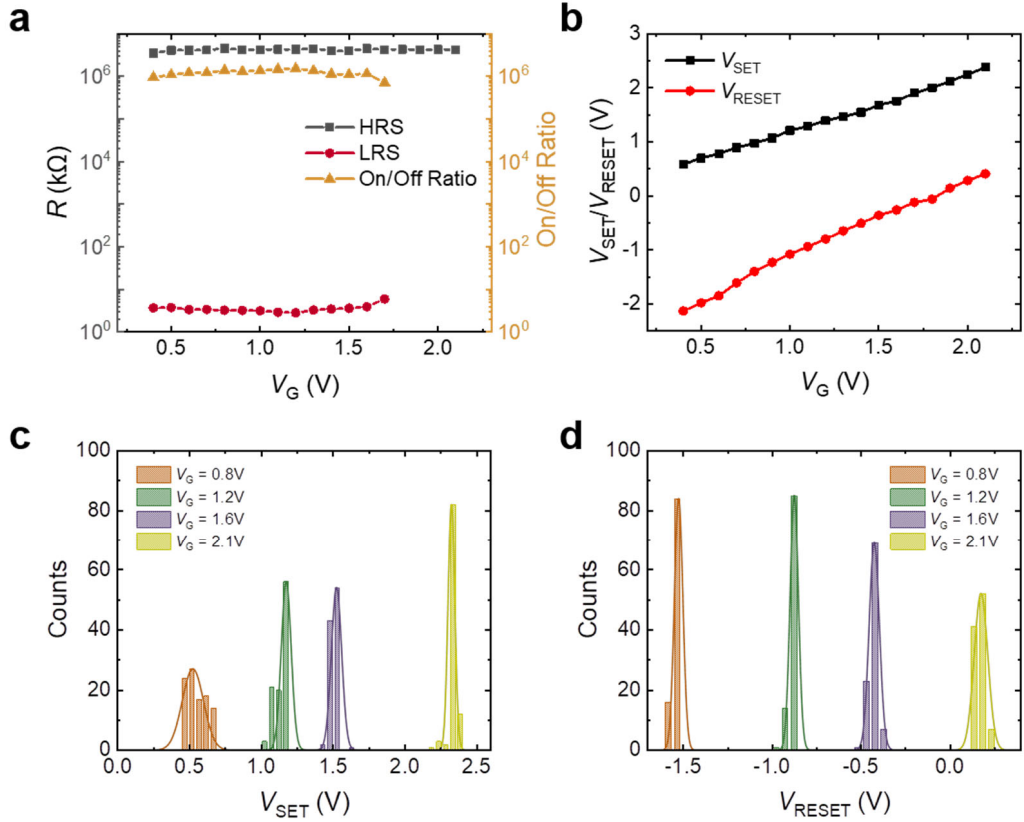

**Figure 3. Tunable and stable switching behaviors in graphene memtransistor.** (a) Resistance of HRS and LRS at various $V_G$. The on/off ratio is observed to be at least $10^6$ and is almost independent of $V_G$. (b) Linear relationship between $V_{SET}$ and $V_{RESET}$ with respect to $V_G$. $V_{SET}$ could be modulated between 0.6V to 2.4V, while the $V_{RESET}$ could be adjusted from -2.1V to 0.4V. (c-d) Histogram for the distribution of $V_{SET}$ and $V_{RESET}$ for 100 cycles. The sharp distribution demonstrates a high degree of precision and stability of switching performance in the graphene memtransistor.

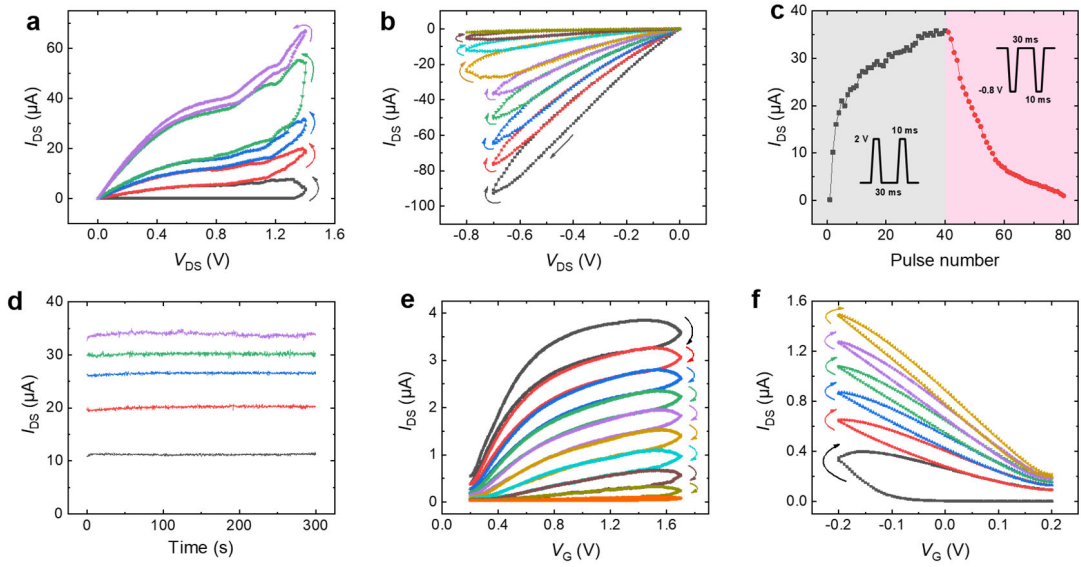

**Figure 4. Synaptic behaviors based on non-volatile graphene memristor.** (a) Continuous increase of $I_{DS}$ by cyclic sweeping of $V_{DS}$ (0 V to 1.4 V) at fixed $V_G$ = 1 V. The change in channel conductance emulates the modification of artificial synaptic weight in respond to the input voltage. The I-V trace of each subsequent upward sweeping overlaps with the previous downward one in the range of $V_{DS}$ < 1.0 V, indicating the high stability of synaptic weight. ( b) Gradual decrease in the conductance of graphene memristor during successive sweeps of negative $V_{DS}$. (c) Incremental increase or decrease in conductance by consecutive potentiating or depressing pulses. The conductance could be incrementally increased or decreased by consecutive potentiating or depressing pulses. (d) Retention properties at different conduction states obtained after applying different numbers of programming pulses with a read voltage of 50 mV. All the data shown in (a-d) are collected with fixed $V_G$ = 1 V. (e-f) Conductance modulation by using the cyclic sweeping of $V_G$ in the configuration of synaptic transistor with fixed $V_{DS}$ = 5 mV. The starting voltage of $V_G$ is set to 0.2 V, where the Fermi level is at the Dirac point of graphene.

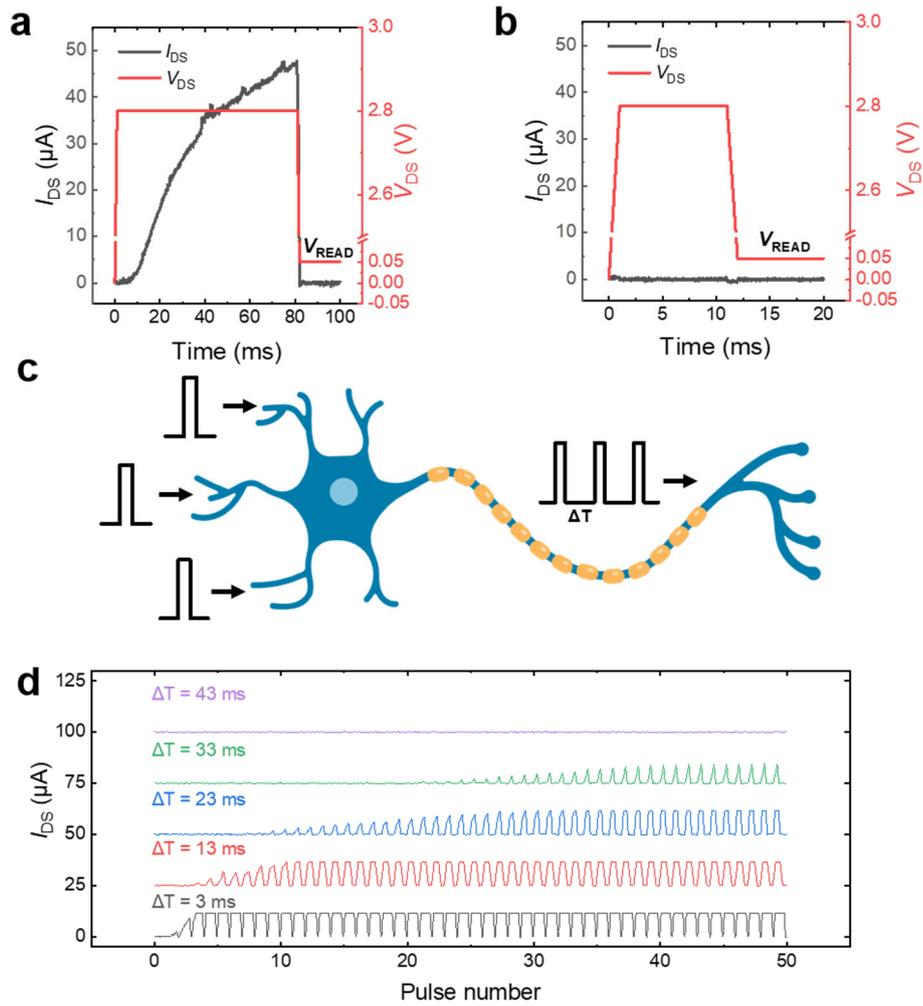

**Figure 5. Neuron behaviors based on the volatile graphene memristor at $V_G$ = 2 V.** (a-b) The characterization of volatile switching behavior under a single voltage pulse. The neuron spikes only if the pulse amplitude and width exceed a certain threshold. A read voltage (0.05V) is applied to detect the conductance after the excitation. (c) Schematic neuron structure and LIF model. (d) Output current modulation in respond to the pulse train with various time intervals (ΔT). ΔT is the time interval between two adjacent pulses as shown in (c). The applied pulse amplitude is 2.8 V with 15 ms width. Each curve is offset by 25 μA for clarity.

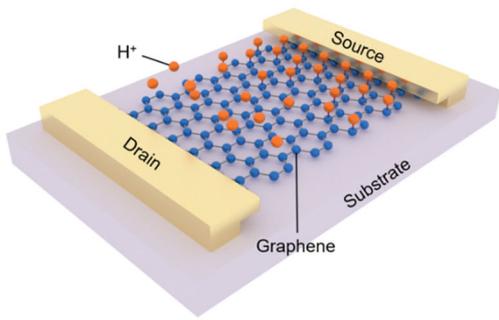 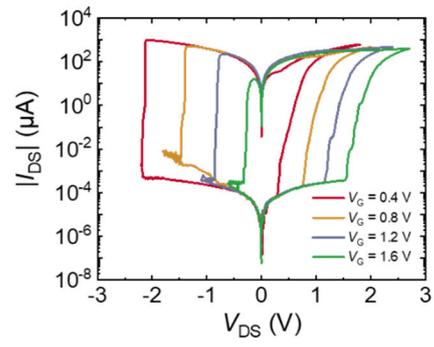
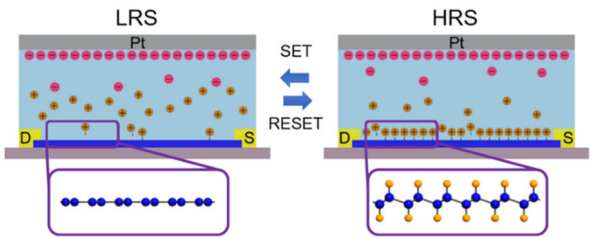 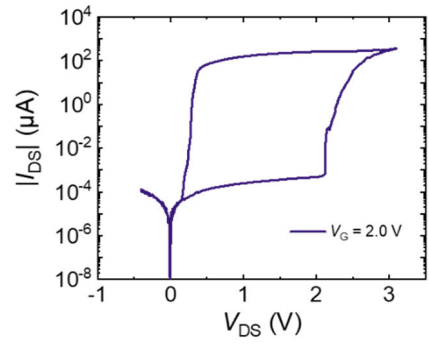

**TOC Graphic**